\documentstyle[12pt]{article}

\font\tenrm=cmr10
\font\tenit=cmti10
\font\elevenbf=cmbx10 scaled\magstep 1
\font\elevenrm=cmr10 scaled\magstep 1
\font\elevenit=cmti10 scaled\magstep 1

\font\ninerm=cmr9

\newcommand{\arcth}{{\rm arcth}\,}
\newcommand{\Tr}{{\rm Tr}\,}
\renewenvironment{thebibliography}[1]
 { \elevenrm
   \begin{list}{\arabic{enumi}.}
    {\usecounter{enumi} \setlength{\parsep}{0pt}
     \setlength{\itemsep}{3pt} \settowidth{\labelwidth}{#1.}
     \sloppy
    }}{\end{list}}

\begin{document}
\hfill NYU-TH-93/09/06
\vskip .01in
\hfill  hep-th/9309114
\begin{center}
\vglue 0.6cm
{\elevenbf \vglue 10pt
OPEN STRINGS IN CONSTANT ELECTRIC AND MAGNETIC FIELDS\footnote{
\ninerm\baselineskip=11pt Talk given at
the conference ``Strings '93'', Berkeley, CA, (May 24-29, 1993) to appear
in the proceedings.}  \\}
\vglue 1.0cm
{\tenrm MASSIMO PORRATI\footnote{
\ninerm\baselineskip=11pt On leave of absence from INFN, Sez. di Pisa, Pisa,
Italy.}\\}
\baselineskip=13pt
{\tenit Department of Physics, New York University, 4 Washington Place\\}
\baselineskip=12pt
{\tenit New York, NY 10003, USA\\}

\vglue 0.8cm
{\tenrm ABSTRACT}

\end{center}

\vglue 0.3cm
{\rightskip=3pc
 \leftskip=3pc
 \tenrm\baselineskip=12pt
 \noindent
Various properties of open strings in external constant E.M. fields are
reviewed.
In particular, the charged-particle pair production rate in an external
electric field is evaluated, and shown to reduce to Schwinger's formula in the
limit of low-intensity fields. Open strings in external magnetic fields are
shown to undergo an infinite number of phase transitions as the strenght of
the field increases.}
\vglue 0.6cm
{\elevenbf\noindent 1. Open Strings in a Constant Electric-Field Background}
\vglue 0.2cm
\baselineskip=14pt
In the presence of an external constant {\em electric} field, quantum field
theory predicts a nonvanishing probability for the creation of charged
particle pairs. The rate of particle pair production can be estimated using
simple semiclassical arguments. Namely, one represents a virtual
charged particle pair in an external electric field of strenght $E$
by a potential well of depth $2m$, where $m$ is the mass
of the charged particle, superimposed to a linear potential
\begin{equation}
V=0,\;\;\; x<0,\;\;\; V=2m-eEx,\;\;\; x\geq 0.
\label{1}
\end{equation}
By denoting with $e$ the charge of the particle and using the standard
semi-classical approximation one would get a rate
$w\sim\exp (-{\cal O}(1) m^2/|eE|)$. This estimate is quite accurate. Indeed,
the {\em exact} rate for a particle of spin $s$ minimally coupled to an
external electric field was found long ago by Schwinger{}${}^1$:
\begin{equation}
w={2s+1\over 8\pi^3} \sum_{k=1}^\infty (-1)^{(2s+1)(k+1)}(eE/ k)^2
\exp(-k\pi m^2/|eE|).
\label{2}
\end{equation}

A natural question arising at this point is whether it is possible
to find an exact
formula for the pair-production rate in string theory. The answer to this
question is in the affirmative, in the case of the open bosonic and
supersymmetric strings{}${}^2$.

Open bosonic strings in an external {\em constant} e.m. field are
exactly soulble, to
lowest order in the string coupling constant. Their world-sheet action reads,
in the gauge $A_\mu=(1/2)F_{\nu\mu}x^\nu$
and for a state of total electric charge $e_1+e_2${}${}^{2,3,4,5}$
\begin{eqnarray}
S&=&-{1\over 2\pi} \int d\sigma d\tau \partial_a X^\mu \partial^a X_\mu +
\nonumber\\
& & +
{1\over 2} e_1 \int d\tau F_{\mu\nu} X^\nu \partial_\tau X^\mu \big|_{\sigma=0}
+{1\over 2} e_1 \int d\tau F_{\mu\nu} X^\nu \partial_\tau X^\mu
\big|_{\sigma=\pi}.
\label{3}
\end{eqnarray}
Here we set $\alpha'=1/2$. The world-sheet coordinates are $\sigma \in [0,\pi]$
and the proper time $\tau$. The 26-dimensional target-space metric is
$\eta_{\mu\nu}=(-1,+1,..,+1)$.
The e.m. field couples only through boundary terms, thus the equations of
motion of all coordinates are the same as in the free-string case
\begin{equation}
\partial_a\partial^a X^\mu=0.
\label{4}
\end{equation}
The only effect of the e.m. field is to modify the boundary conditions obeyed
by the $X^\mu$. Notice that the target-space metric remains flat to lowest
order in the string coupling constant $g=e^\phi$.
The back reaction due to the presence of a non-zero stress-energy tensor is
in fact of order $g$ with respect to the tree-level metric.
When the external field is purely electrical one can choose a coordinate system
in which only the component $F_{01}$ of $F_{\mu\nu}$ is non-vanishing. By
introducing standard light-cone coordinates $X^{\pm}= (X^0\pm X^1)/\sqrt{2}$
the boundary conditions read
\begin{eqnarray}
\partial_\sigma X^\pm &=&\pm \beta_1 \partial_\tau X^\pm,\;\;\;\; \sigma=0,
\nonumber\\
\partial_\sigma X^\pm &=&\mp \beta_2 \partial_\tau X^\pm,\;\;\;\; \sigma=\pi.
\label{5}
\end{eqnarray}
In Eq.~(\ref{5}) $\beta_i\equiv \pi e_iE$, $E=F_{01}$.
The boundary conditions and mode expansion for the
transverse coordinates $X^i$ $i=2,..,25$ are the same as for the free open
string. Equations~(\ref{4},\ref{5})
imply instead the following mode expansion
for the coordinates $X^\pm$
\begin{eqnarray}
X^\pm &=& x^\pm + ia^\pm_0 \phi^\pm_0 +i \big( \sum_{n=1}^\infty a_n^\pm
\phi^\pm_n(\sigma,\tau) -h.c. \big), \nonumber \\
\phi_n^\pm &=& (n\mp i\epsilon)^{-1/2} e^{-i(n\mp i\epsilon)\tau}
\cos[(n\mp i\epsilon)\sigma \pm \arcth \beta_1], \nonumber\\
(a_0^\pm)&=& \pm i a_0^\pm,\;\;\; \epsilon ={1\over \pi}(\arcth \beta_1 +
\arcth \beta_2 ).
\label{6}
\end{eqnarray}
Notice that for small electric fields, that is for fields
$e_iE\ll \alpha'^{-1}$, $\epsilon\approx 2\alpha'(e_1+e_2)E$.
The main effect of the
non-vanishing electric field is an {\em imaginary}
shift in the free-string frequencies
$n \rightarrow n \mp i\epsilon$. Moreover, similarly to the case of a particle
in an external magnetic field, the center-of-mass coordinates $x^\pm$ do not
commute, but obey instead the equation
\begin{equation}
[x^+,x^-]={-i\pi \over \beta_1 +\beta_2}.
\label{7}
\end{equation}

By using the above mode expansion one easily finds the Virasoro operators
$L_n$. They can be written as a sum of a ``transverse'' component
$L^{\perp}_n$, involving only the coordinates $X^2,..,X^{25}$,
and identical to the free-string one, and a ``longitudinal'' one,
$L^{\parallel}_n$. The component $L^{\parallel}_0$ reads, for $\epsilon > 0$,
\begin{equation}
L_0^{\parallel} = - \sum_{n=0}^\infty (n+i\epsilon) (a^+_n)^* a^-_n -
\sum_{n=1}^\infty (n-i\epsilon)(a^-_n)^*a_n^+ +{1\over 2}
i\epsilon(1-i\epsilon).
\label{8}
\end{equation}
Besides the imaginary shift in frequencies, the main difference with the
free-string case is the shift in the vacuum energy. This shift can be
determined either by spectral flow (with a twist
$\theta=i\epsilon $){}${}^2$,
or by imposing that the Virasoro algebra closes in the standard form
$[L_m,L_n]=(n-m)L_{n+m} +1/12cm(m^2-1)${}${}^{3,4}$.

The vacuum to vacuum transition amplitude $\langle 0|\exp -iTH|0\rangle $
is given in terms of the free-energy density $F$
by a standard field-theory argument
as $\exp(-iTVF)$. Thus the total pair-production rate $w$, equal to the
probability of vacuum decay, is $w=-2\Im F$.

In open string theory the free-energy density
$F$ is given, at one loop, by a sum of
four terms arising from the four different geometries open strings can
propagate on: the torus, Klein bottle, annulus, and M\"obius strip
\begin{equation}
-iTVF= {1\over 2}{\cal T} + {1\over 2}{\cal K} + {1\over 2}{\cal A}
  + {1\over 2}{\cal M}
\label{9}
\end{equation}
The first two geometries correspond to closed string states: they do not
depend on the end-point charges $e_1$ and $e_2$, since the torus and the Klein
bottle have no boundaries, thus they do not give rise to
terms contributing to $\Im F$.

The annulus contribution reads
\begin{equation}
{\cal A}={1\over 2} \lim_{\delta\rightarrow 0}\int_\delta^\infty dt t^{-1}
\Tr e^{-\pi t(L_0 -1)} \approx {1\over 2} \Tr \log (L_0-1).
\label{10}
\end{equation}
The trace in Eq.~(\ref{10}) is taken over all variables (oscillators, momenta
etc.) and over all possible charge sectors of the open string. These charge
sectors, in a consistent model, are determined by the embedding of
$U(1)_{e.m.}$ into the open string gauge group.
The integration variable $t$ is the (real) modular parameter of the annulus.

The M\"obius strip has only one boundary, and thus contributes only to sectors
where $e_1=e_2$. Its contribution to the free-energy is
\begin{equation}
{\cal M}= \pm {1\over 2} \lim_{\delta\rightarrow 0}
\int_\delta^\infty dt t^{-1} \Tr {\cal P} e^{-\pi t(L_0-1)}.
\label{11}
\end{equation}
The operator ${\cal P}$ implements the change of orientation of the open
string $\sigma \rightarrow \pi-\sigma$.
The M\"obius-strip contribution, together with the Klein-bottle one,
is needed in order to cancel the small-$t$ divergences arising already in
the free-string annulus integral.
Only after annulus, Klein-bottle and M\"obius-strip contributions are added,
and only for a specific choice of the gauge group, do the free-string
small-$t$ divergencies cancel. The correct gauge group turns out to be
$SO(8192)$ for the bosonic string, and $SO(32)$ for the open
superstring${}^6$.

Cancellation of divergencies in the presence of an external field
requires to take into proper account the back reaction on the target-space
metric induced by the presence of the background field.
However, since we are only
interested in the imaginary part of the free-energy, which receives no
contribution form the small-$t$ region of integration, we can safely
ignore this complication.

The annulus contribution to the free-energy can be evaluated
straightforwardly, since the Virasoro operators are sums of free oscillators
\begin{equation}
A=\sum_{e_1,e_2\in Q} \lim_{\delta\rightarrow 0}
VT {|\beta_1 +\beta_2|\over (2\pi)^2}\int_\delta^\infty dt t^{-1} \int
{d^{24}p\over (2\pi)^{24}} e^{-\pi t p^2/2} T^{\perp} T^{\parallel}.
\label{12}
\end{equation}
In this equation $Q$ denotes the set of boundary charges determined by the
embedding of $U(1)_{e.m.}$ into the open-string gauge group. $T^{\perp}$ and
$T^{\parallel}$ arise from taking the trace over transverse and longitudinal
oscillators, respectively. The integration over transverse momenta is
standard. The normalization factor in front of the integral is the only place
where the $\beta_i$ enter explicitly, instead of $\epsilon$, and it is
determined by noticing that the commutation relation~(\ref{7}) implies that the
$x^\pm$ phase-space integration measure is
\begin{equation}
 {|\beta_1+\beta_2|\over 2\pi^2} dx^+dx^-.
\label{13}
\end{equation}
Obviously $T^{\perp}$ has the same form as for the free open
string, since it involves only transverse oscillators
\begin{equation}
T^{\perp}= \eta(it/2)^{-24}\eta(it/2)^2.
\label{14}
\end{equation}
Notice the contribution $\eta^2$ coming from the coordinate-reparametrization
ghosts.

$T^\parallel$ depends on the external electric field and reads
\begin{equation}
T^\parallel = e^{-\pi t(\epsilon^2/2-1/12)}\left[2i
\sin(\pi|\epsilon|t/2)\prod_{n=1}^\infty \left|1-e^{-\pi t(n
+i\epsilon)}\right|^2\right]^{-1}.
\label{15}
\end{equation}
This quantity is imaginary, as imposed by equation~(\ref{9}).
The free-energy $F$ though, has a nonzero imaginary part, arising from the
integration in $t$. Indeed, $T^{\parallel}$ has the form
\begin{equation}
T^\parallel= [2i\sin(\pi|\epsilon|t/2)]^{-1} f(t),
\label{16}
\end{equation}
with $f(t)$ regular for $t>0$, and thus it has simple poles
for positive $t$, located
at $t=2k/|\epsilon|$, $k$ integer.
When integrating in $t$ one has to deform the
contour of integration so as to avoid these poles. The correct prescription
on the contour is, as expected, the one allowing for the analytic continuation
$t\rightarrow it$. Thus, $\Im F$ reduces to a sum over residues at the
poles, according to Cauchy's theorem
\begin{equation}
{\cal A}=\sum_{e_1,e_2\in Q} VT{\beta_1 +\beta_2 \over  2
(2\pi)^{26}\epsilon }
\sum_{k=1}^\infty (-1)^{k+1} e^{-\pi|\epsilon|k}(|\epsilon|/k)^{13}
\eta(ik/|\epsilon|)^{-24}.
\label{17}
\end{equation}
This formula can be recast in the following simple form
\begin{equation}
{\cal A}= \sum_{e_1,e_2\in Q} VT{\beta_1 +\beta_2 \over 2\epsilon }
\sum_{k=1}^\infty (-1)^{k+1}Z(2k/|\epsilon|),
\label{18}
\end{equation}
where $Z(t)$ is the {\em free-string} partition function on the annulus.
Equation~(\ref{18}) holds in any compactification that
does not involve the $X^\pm$ coordinates. Notice also that the term under the
$Q$-summation sign in Eq.~(\ref{18}) gives
the complete pair-production rate for open-string states with $e_1\neq e_2$,
since in this case $F$ receives no contribution from the M\"obius strip.
To compare Eq.~(\ref{18}) with Schwinger's result~(\ref{2}) we must
compactify the bosonic
string to four dimensions
and recall that four distinct open-string sectors contribute to the
production rate of a particle pair of given charge $e=e_1+e_2$. These are the
sectors with boundary charges $(e_1,e_2)$, $(e_2,e_1)$, $(-e_1,-e_2)$, and
$(-e_2,-e_1)$. By denoting with $M_S$ the mass of an open-string state $S$
and after a simple computation one finds
\begin{equation}
w=\sum_S\sum_{k=1}^\infty (-1)^{k+1} {\beta_1+\beta_2\over
8\pi^4\epsilon}(|\epsilon|/k)^2e^{-k\pi(M_S^2+ \epsilon^2)/|\epsilon|}.
\label{19}
\end{equation}
When $\beta_i\ll 1$, $\epsilon\rightarrow (e_1+e_2)E $, and Eq.~(\ref{19})
reduces to Eq.~(\ref{2}).

When the two boundary charges are equal, only two open-string sectors
contribute to the production rate of a given particle pair: $(e,e)$ and
$(-e,-e)$. The annulus amplitude in Eq.~(\ref{17}), therefore, would reduce to
one-half of Schwinger's result. This discrepancy is resolved by taking into
account the contribution of the M\"obius strip to equation~(\ref{9}).

For a given state $S$
this contribution is equal to annulus one, up to a sign. This sign is positive
or negative according to whether the state $S$ is even or odd under
${\cal P}$ (defined in Eq.~(\ref{11})). The sum of the M\"obius and annulus
contribution thus reduces to Schwinger's formula for all states which are not
eliminated from the physical spectrum by the M\"obius projection.

An interesting property of Eqs.~(\ref{17},\ref{18}) is that they diverge for
a (finite) critical value of the electric field, namely when the force applied
by the electric field on either of the boundary charges equals the string
tension
\begin{equation}
\min_i \left | e_i E\right| =(2\pi\alpha')^{-1}.
\label{20}
\end{equation}
This instability arises already at the classical level{}${}^4$.
It is interesting to notice that when $(e_1+e_2)\rightarrow 0$, but $e_1\neq
0$, the annulus partition function ${\cal A}$
does not reduce to the free-string one. In
fact, by setting $\beta_1=-\beta_2+\Delta$, and taking the limit
$\Delta\rightarrow 0$, one finds $\epsilon = \Delta/(1-\beta_1^2)\pi + {\cal
O}(\Delta^2)$. Using Eqs.~(\ref{12},\ref{14},\ref{15}) one then finds that
the neutral-string annulus amplitude ${\cal A}_{neutral}$ and the free-string
one ${\cal A}_{free}$ are related by
\begin{equation}
{\cal A}_{neutral} = (1-\beta_1^2){\cal A}_{free}.
\label{20'}
\end{equation}
This formula agrees with ref.{}${}^3$.

Open superstrings in an external field can be solved exactly at lowest
order in the string coupling constant by using the same procedure already
outlined in the bosonic-string case.
The world-sheet action of the critical 10-dimensional superstring reads
\begin{eqnarray}
S &=& S_{bosonic} +{i\over 2\pi} \int d\sigma d\tau \bar{\psi}^\mu \rho^\alpha
\partial_\alpha \psi_\mu + \nonumber \\
& & +{i\over 4} \left.
e_1\int d\tau F_{\mu\nu} \bar{\psi}^\mu \rho^0 \psi^\nu \right|_{\sigma=0}
+   {i\over 4} \left.
e_2\int d\tau F_{\mu\nu} \bar{\psi}^\mu \rho^0 \psi^\nu \right|_{\sigma=\pi}.
\label{21}
\end{eqnarray}
The $\rho^i$ are the 2-dimensional Dirac matrices.

The appropriate boundary conditions on the 2-dimensional
bosons $X^\pm$ are given in
Eq.~(\ref{5}). The boundary conditions on their fermionic partners $\psi^\pm$
are
\begin{eqnarray}
(1\mp \beta_1)\psi^\pm_R &=& (1\pm \beta_1) \psi_L^\pm, \;\;\; \sigma=0,
\nonumber \\
(1\pm \beta_2)\psi^\pm_R &=& -(-1)^a (1\mp \beta_2)\psi_L^\pm, \;\;\;
\sigma=\pi.
\label{22}
\end{eqnarray}
In this equation the subindices $R$, $L$ denote the 2-dim. handedness of the
fermions. The variable $a$ is equal to 0 when the fermions are given
anti-periodic (Neveu-Schwarz) boundary conditions and to 1 when they are given
periodic (Ramond) boundary conditions.
The transverse fermionic coordinates obey standard free-superstring boundary
conditions.

All fermionic coordinates obey free equations of motion which, together with
Eq.~(\ref{22}), completely determine the fermion mode expansion.
Canonical quantization gives rise to the Virasoro operators
$L_n=L_n^F + L_n^B$. The bosonic contribution to these operators is denoted
by $L_n^B$ and is the same as in the bosonic open string. The fermionic
contribution $L_n^F$ can be further decomposed into a sum of two terms:
$L_n^{F\perp}$, containing only transverse oscillators, and identical to the
corresponding operator for a free open superstring, and $L_n^{F\parallel}$,
depending only on the fermions $\psi^\pm$.
By introducing the canonical anti-commuting operators
$d_n^\mu$ obeying $\{ d^\mu_n,d^\nu_m\}=\eta^{\mu\nu} \delta_{n+m,0}$
one finds, in particular,
\begin{equation}
L_0^{F\parallel}= \sum_{n\in Z+1/2 +a/2} -(n+i\epsilon)
:d^-_{-n}d^+_n: + {a\over 8}
-{i\epsilon\over 2}(a-i\epsilon),\;\;\; \epsilon >0.
\label{23}
\end{equation}
The shift in the vacuum energy can be determined by spectral flow, once a
normal-ordering prescription for $d^\pm_{k}$ is given, or by
closure of the Virasoro algebra in exact analogy with the bosonic case. Notice
that the shift in the Ramond-sector vacuum energy exactly cancels between
bosons and fermions (cfr. Eq.~(\ref{8})).

The annulus amplitude now reads
\begin{equation}
{\cal A} =\sum_{e_1,e_2\in Q}\sum_{a,b} \lim_{\delta\rightarrow 0}
VT {|\beta_1 +\beta_2|\over (2\pi)^2}
C\left[ \begin{array}{c} a\\ b \end{array} \right]
\int_\delta^\infty dt t^{-1} \Tr_a \left[ (-1)^{bF} e^{-\pi t(L_0 -1)}
\right].
\label{24}
\end{equation}
$F$ is the fermion-number operator commuting with all world-sheet bosons and
anti-commuting with the world-sheet fermions. The values of the parameters
$a$ and $b$ are 0 or 1, and $a$ is defined as before to be 0 in the
Neveu-Schwarz sector and 1 in the Ramond sector. The weights in Eq.~(\ref{24})
are chosen so as to implement space-time supersymmetry in 10
dimensions
\begin{equation}
C\left [\begin{array}{c} 0\\ 0 \end{array} \right]=
-C\left [\begin{array}{c} 1\\ 0 \end{array} \right]=
-C\left [\begin{array}{c} 0\\ 1 \end{array} \right]=
\pm C\left [\begin{array}{c} 1\\ 1 \end{array} \right]={1\over 2}.
\label{25}
\end{equation}
the trace in Eq.~(\ref{24}) factorizes into the product of a purely bosonic
contribution, which is evaluated using Eqs.~(\ref{12},\ref{14},\ref{15}),
and a fermionic one, which reads
\begin{eqnarray}
\Tr_a (-1)^{bF} e^{-\pi t L_0^F} &=& T_F^{ghost}T_F^\perp T_F^\parallel,
\nonumber\\
T_F^{ghost} &=& \Tr_a (-1)^{bF} e^{-\pi t L_0^{F\; ghost}} =
\eta(it/2)/\Theta\left[ \begin{array}{c} a\\ b \end{array}
\right] (0| it/2), \;\;\; a+b\neq 2, \nonumber \\
T_F^\perp &=& \Tr_a (-1)^{bF} e^{-\pi t L_0^{F\perp}} =
\left\{\Theta\left[ \begin{array}{c} a\\ b \end{array}
\right] (0| it/2)/\eta(it/2)\right\}^3, \nonumber \\
T^\parallel &=& \Tr_a (-1)^{bF} e^{-\pi t L_0^{F\parallel}} =
\Theta\left[ \begin{array}{c} a-2i|\epsilon|
\\ b \end{array}\right] (0| it/2)/\eta(it/2).
\label{26}
\end{eqnarray}
The only term depending on the external electric field in Eq.~(\ref{26})
is, as expected, $T_F^\parallel$. Here the conventions on theta functions with
characteristics are as in{}${}^2$.
The contribution $T_F^{ghost}$, arising
from the world-sheet supersymmetry ghosts,
takes a special form in the presence
of a fermionic zero-mode (when $a+b=2$). We need not to consider it because the
contribution of the $a+b=2$ sector to ${\cal A}$ gets cancelled by
$T^\perp_F$, since the theta function with $a=b=1$ vanishes identically.

It is immediate to notice that the poles in the $t$-integration arise only
from the trace over bosons. Thus, the annulus amplitude can be evaluated
exactly as in the bosonic case. The residue at the pole $t=2k/|\epsilon|$
is given by a bosonic term times the fermionic term
\begin{equation}
T_F(t=2k/|\epsilon|)= (-1)^{ak}e^{\pi k|\epsilon|}
\left\{\Theta\left[ \begin{array}{c} a\\ b \end{array}
\right] (0| ik/|\epsilon|)/\eta(ik/|\epsilon|) \right\}^4.
\label{27}
\end{equation}
Notice that this term contributes an extra $(-1)^{k}$ to the sum over residues
when $a=1$, that is for space-time fermions, in analogy with Schwinger's
formula Eq.~(\ref{2}). After a few manipulations, analogous to the ones
performed in the bosonic case, one arrives at the following formula for the
vacuum decay rate when $10-D$ dimensions are compactified
\begin{equation}
w= {1\over (2\pi)^{D-1}}
\sum_{S} {\beta_1 +\beta_2 \over \pi \epsilon } \sum_{k=1}^\infty
(-1)^{(k+1)(a_S +1)} (|\epsilon|/k)^{D/2} e^{-\pi kM_S^2/|\epsilon|}.
\label{28}
\end{equation}
Here the variable $S$ labels the string states surviving the GSO and M\"obius
projections, $a_S=0$ for states in the Neveu-Schwarz sector and
$a_S=1$ for Ramond-sector states. In $D=4$ and for small fields this
expression reduces to Eq.~(\ref{2}). Notice that in Eq.~(\ref{28}) the
production rates for each particle species diverge at $\beta_i=1$. This
behavior is different from the bosonic-string one, where the production rates
{\em vanished} species by species, but their sum diverged, thanks to the
stronger divergence of the statistical sum over all string states.

Finally, one may notice that equation~(\ref{5})
interpolates between free-string (Neumann) boundary conditions, at $\beta_i=0$,
and reflecting boundary conditions at $\beta_i=1$. This latter case
corresponds to the critical value of the electric field given in
Eq.~(\ref{20}). At $\beta_i=1$, thus, the world-sheet dynamics of the bosonic
string becomes formally equivalent to that of a black-hole solution of
2-dimensional gravity coupled to 24 free massless scalars{}${}^7$ (see also
K. Schoutens's and E. Verlinde's contributions to these proceedings).
\vglue 0.6cm
{\elevenbf\noindent 2. Open Strings in a Constant Magnetic-Field Background}
\vglue 0.2cm
Open strings in an external purely magnetic field exhibit new interesting
features already at tree-level{}${}^8$.
Let us examine at first the bosonic string.
In this case
by choosing a coordinate system in which the only non-vanishing
component of the magnetic field is $H\equiv H_{12}$, the Virasoro operator
$L_0$ reads{}${}^{3,8}$
\begin{equation}
L_0=
 (2b_0^\dagger b_0+1){|h|\over 2} -{1\over 2} h^2
-|h|\sum_{k=1}^\infty(a^\dagger_ka_k-b^\dagger_k b_k) + L_0^{free}.
\label{30}
\end{equation}
Here $L_0^{free}$ denotes the free-string Virasoro operator of all the
transverse coordinates $X^0,X^3,...,X^{(D-1)}$, and we
introduced the complexified creation and annihilation operators
\begin{equation}
\sqrt{2} a_k=\alpha^1_k +i\,{\rm sign}\,(h)\alpha^2_k,\;\;\;
\sqrt{2} b_k=\alpha^1_k -i\,{\rm sign}\,(h)\alpha^2_k.
\label{31}
\end{equation}
Equation~(\ref{30}) differs from the free string expression by terms
involving
\begin{equation}
h={1\over \pi} (\arctan 2\alpha' e_1\pi H + \arctan 2\alpha' e_2 \pi H).
\label{32}
\end{equation}
Notice in particular the presence of a magnetic-dipole coupling, non-linear in
the magnetic-field strenght
\begin{equation}
|h|\sum_{k=1}^\infty (a_k^\dagger a_k -b_k^\dagger b_k) =h S_{12},
\label{33}
\end{equation}
where $S_{\mu\nu}$ is the target-space spin operator. In particular, in four
dimensions, $S_{12}$ is the spin component along the magnetic field. Since, in
the low-field limit ($\alpha'eH\ll 1$)
$h\rightarrow 2\alpha'(e_1+e_2)H$, it is apparent that open string
states have all a gyromagnetic ratio $g=2${}${}^{9,10}$. Moreover, in $D=4$
and in the
low-field limit, Eq.~(\ref{30}) itself reduces to the well-known
field-theoretical
formula giving the energy levels of a particle of mass $M$, spin $\vec{S}$,
and gyromagnetic ratio $g$ in an external magnetic field
(see for instance{}${}^{11}$)
\begin{equation}
E^2=(2n+1)eH -g_S e \vec{H}\cdot \vec{S} + p_3^2 + M^2.
\label{33'}
\end{equation}
To arrive at this formula,  it suffices to recall that the wave equation
of a string state $\Psi$ is
\begin{equation}
(L_0-1)\Psi\equiv \alpha'[E^2-(p^0)^2]\Psi=0,
\label{33''}
\end{equation}
and to notice that the eigenvalues $n$ of  $b^\dagger_0 b_0$ are the
Landau levels.
The Virasoro operator $L_0$ possesses the usual tachionic mode of the free
bosonic string, but it also shows additional tachionic modes when $H\neq 0$.
By direct examination of Eq.~(\ref{30}) one may see that
only the highest-helicity states belonging to the first (``parent'') Regge
trajectory can become tachyonic.
By denoting with $|0\rangle$ the Fock vacuum for the string
oscillators $a_k$, $b_k$ and $\alpha_k$, these states read
\begin{equation}
\Psi(m)=\left( a_{1}^\dagger\right)^m|0\rangle .
\label{34}
\end{equation}
The value of $\alpha'E^2$ on them is
\begin{equation}
(\alpha'E^2)\Psi(m)=
\left[ {|h|\over 2} -{h^2\over 2} + (1-|h|)m -1\right]\Psi(m).
\label{35}
\end{equation}
For fixed value of the magnetic field, and thus of $|h|$, $\alpha'E^2$
is negative on all states with
\begin{equation}
m < {1+ |h|(|h|-1)/2\over 1-|h|}.
\label{36}
\end{equation}
In particular, the highest helicity of the open-string massless vector ($m=1$)
is always tachionic when $H\neq 0$. This instability is well-known
in Yang-Mills theory{}${}^{12}$. The $m=0$ instability  of open bosonic
strings was also noticed in ref.{}${}^3$. There is an infinite number of
states given by Eq.~(\ref{35}), correspondingly, as the intensity of the
external magnetic field increases, open strings may
undergo an infinite number of
phase transitions. As $H$ approaches infinity $h\rightarrow 1$. In this limit
all states $\Psi(m)$ would become tachionic, in the standard supersting
vacuum, with a common (negative) square energy $\alpha'E^2=-1$.

The meaning of these phase transitions is that the vacuum where the VEVs of
the $\Psi(m)$ vanish is unstable, when the magnetic field passes
the threshold value given by Eq.~(\ref{36}): in the stable vacuum
$\Psi(m)$-condensates should appear (cfr.{}${}^{12,13}$).

The previous analysis can be straightforwardly extended to the open
superstring.

In the superstring case one must introduce fermionic oscillators, either in
the Ramond or Neveu-Schwarz sector, besides the bosonic ones. The fermionic
oscillators can be complexified in the same way as the bosonic coordinates
\begin{equation}
\sqrt{2} d_k= d^1_k +i\,{\rm sign}\,(h) d^2_k,\;\;\;
\sqrt{2} \tilde{d}_k= d^1_k -i\,{\rm sign}\,(h) d^2_k
\label{37}
\end{equation}

Using the same notations as before one finds that the Virasoro operator $L_0$
takes the following form in the Ramond sector
\begin{equation}
L_0= (2n+1){|h|\over 2} + |h|d_0^\dagger d_0 -{|h|\over 2}
 -|h|\sum_{k=1}^\infty\left( d^\dagger_k d_k- \tilde{d}^\dagger_k \tilde{d}_k
+a^\dagger_k a_k -b^\dagger_k b_k \right) + L_0^{free}.
\label{38}
\end{equation}
Recalling that the wave equation in the Ramond sector takes the form
$L_0\Psi=0$ one easily finds that all Ramond states have positive square
energy. This result follows from the inequality $h\leq 1$.
This inequality is due
to the fact that in string theory the magnetic-dipole coupling
is non-linear (see Eq.~(\ref{32})). The field-theoretical formula~(\ref{33'})
with $g=2$ would instead give instabilities for {\em any} spin larger than
1/2.

The Virasoro operator in the Neveu-Schwarz sector reads
\begin{equation}
L_0= (2n+1){|h|\over 2}
 -|h|\sum_{k=1/2}^\infty
\left( d^\dagger_k d_k- \tilde{d}^\dagger_k \tilde{d}_k
+a^\dagger_k a_k -b^\dagger_k b_k \right) + L_0^{free}.
\label{39}
\end{equation}
The Neveu-Schwarz wave equation is $(L_0-1/2)\Psi=0$, this implies that there
exist Neveu-Schwarz physical states (i.e. states surviving the GSO projection)
with $\alpha'E^2<0$, when $H$ is sufficiently large. A brief analysis of
Eq.~(\ref{39}) shows that these states have the form
\begin{equation}
(a_{1}^\dagger)^md^\dagger_{1/2}|0\rangle_{NS},
\label{40}
\end{equation}
and that they become tachionic when
\begin{equation}
m < {|h|\over 2(1-|h|)}.
\label{41}
\end{equation}
They are, again, highest-helicity states belonging to the parent Regge
trajectory of the Neveu-Schwarz sector.  When $h\rightarrow 1$ the
negative square energy of all these states becomes $\alpha'E^2=-1/2$.

An interesting and difficult problem still to be solved is to find an ansatz
for the  scalar potential of the fields given in Eqs.~(\ref{34},\ref{40}).
Knowledge of this potential would allow a detailed study of high magnetic field
phase
transitions in string theory.
\vglue 0.6cm
{\elevenbf\noindent References \hfil}
\vglue 0.4cm

\end{document}